\documentclass{article}

\makeatletter

\@addtoreset{equation}{section}
\makeatother

\usepackage{amsmath,amssymb}
\usepackage{cite}
\usepackage{graphicx}
\usepackage{hyperref}

\newcommand {\beq}{\begin{eqnarray}}
\newcommand {\eeq}{\end{eqnarray}}

\newcommand{\coh}{\operatorname{coh}}
\newcommand{\module}{\operatorname{mod}}

\newcommand{\bC}{\ensuremath{\mathbb{C}}}

\newcommand{\bN}{\ensuremath{\mathbb{N}}}

\newcommand{\bR}{\ensuremath{\mathbb{R}}}

\newcommand{\bT}{\ensuremath{\mathbb{T}}}

\newcommand{\bZ}{\ensuremath{\mathbb{Z}}}

\newcommand{\scM}{\ensuremath{\mathcal{M}}}

\begin{document}
\baselineskip=18pt  

\begin{titlepage}

\setcounter{page}{0}

\renewcommand{\thefootnote}{\fnsymbol{footnote}}

\begin{flushright}
CALT-68-2706\\
IPMU-08-0087\\
UT-08-30\\
\end{flushright}

\vskip 1.35cm

\begin{center}
{\LARGE \bf
Crystal Melting and Toric Calabi-Yau Manifolds}

\vskip 1.5cm 

{\large
Hirosi Ooguri$^{1,2}$ and Masahito Yamazaki$^{1,2,3}$
}

\vskip 0.8cm

{\large \sl 
$^1$ California Institute of Technology, 452-48, Pasadena, CA 91125, USA\\
\medskip
$^2$Institute for the Physics and Mathematics of the Universe,\\
University of Tokyo, Kashiwa, Chiba 277-8586, Japan\\
\medskip
$^3$Department of Physics, University of Tokyo, \\
Hongo 7-3-1, Tokyo 113-0033,
Japan\\
}

\end{center}

\vspace{1.8cm}

\centerline{{\large \bf Abstract}}
\bigskip
\noindent
We construct a statistical model of crystal melting 
to count BPS bound states of D0 and D2 branes on a single 
D6 brane wrapping an arbitrary toric Calabi-Yau threefold.
The three-dimensional crystalline structure is determined by
the quiver diagram and the brane tiling which characterize 
the low energy effective theory of D branes. The crystal 
is composed of atoms of different colors, each of which 
corresponds to a node of the quiver diagram, and the chemical 
bond is dictated by the arrows of the quiver diagram. BPS states 
are constructed by removing atoms from the crystal. 
This generalizes the earlier results on the BPS state 
counting to an arbitrary non-compact toric Calabi-Yau 
manifold. We point out that a proper understanding
of the relation between the topological string theory 
and the crystal melting involves the wall crossing in
 the Donaldson-Thomas theory. 

\end{titlepage}
\setcounter{page}{1} 

\section{Introduction}

In type IIA superstring theory, supersymmetric bound states of D branes 
wrapping holomorphic cycles on a Calabi-Yau manifold give 
rise to BPS particles in four dimensions. In the past few
years, remarkable connections have been found between 
the counting of such bound states and the topological string
theory:

\medskip
\noindent
(1) When the D brane charges are such that bound 
states become large black holes with smooth event 
horizons, the OSV conjecture \cite{OSV} states that the generating 
function $Z_{{\rm BH}}$ of a suitable index for black hole 
microstates is equal to the absolute value squared of the 
topological string partition function $Z_{{\rm top}}$,
\beq Z_{{\rm BH}}=\left| Z_{{\rm top}}\right|^2, \eeq
to all orders in the string coupling expansion. 

\medskip
\noindent
(2) When there is a single D6 brane 
with D0 and D2 branes bound on it, it has been proposed \cite{INOV}
that the bound states are counted by the Donaldson-Thomas 
invariants \cite{DT,Thomas} of the moduli space of ideal sheaves 
on the D6 brane. For a non-compact toric Calabi-Yau manifold, the 
Donaldson-Thomas invariants are related to the topological string partition 
function \cite{ORV,INOV,MNOP1} using
the topological vertex construction \cite{AKMV}. Recently the 
connection between the topological string theory and the Donaldson-Thomas
theory for toric Calabi-Yau manifolds was proven mathematically 
in \cite{MOOP}. Given the conjectural relation between
the counting of D brane bound states and the Donaldson-Thomas 
theory, it is natural to expect the relation, 
\beq Z_{{\rm BH}}=Z_{{\rm top}}. \label{topstring} \eeq

\medskip

The purpose of this paper is to understand the case (2) better.
We start with the 
left-hand side of the relation, namely the counting of BPS states. 
Recently, the non-commutative 
version of the Donaldson-Thomas theory is formulated by 
Szendr\"oi \cite{Szendroi} for the conifold 
and by Mozgovoy and Reineke \cite{MR} for general toric 
Calabi-Yau manifolds.\footnote{See \cite{Young1,Young2,NN,Nagao,JM,CJ} 
for further developments.}
In this paper, we will establish a direct connection between
the non-commutative Donaldson-Thomas theory and 
the counting of BPS bound states of D0 and D2 branes on 
a single D6 brane. Using this correspondence, we will
find a statistical model of crystal melting which counts
the BPS states. 

The crystal melting description has been found
earlier in the topological string theory 
on the right-hand side of \eqref{topstring}. 
It was shown in \cite{ORV, INOV} that 
the topological string partition function on 
$\bC^3$, the simplest toric Calabi-Yau manifold,
and the topological vertex can be expressed 
as sums of three-dimensional Young diagrams,
which can be regarded as complements of molten
crystals with the cubic lattice structure.\footnote{
See \cite{foamone,foamtwo,foamthree,foamfour,Sulkowski,foamfive,Jafferis,foamseven} 
for further developments.} Since 
the topological vertex can be used to compute
the topological string partition function for
a general non-compact toric Calabi-Yau manifold,
it is natural to expect that a crystal melting 
description exists for any such manifold. To our 
knowledge, however, this idea has not been made
explicit. The crystal melting model defined in 
this paper appears to be different from the one 
suggested by the topological vertex construction.

The low energy effective theory of D0 and D2 branes
bound on a single D6 brane is a one-dimensional
supersymmetric gauge theory, which is a dimensional reduction
of an ${\cal N}=1$ gauge theory in four dimensions.
The field content of the gauge theory is encoded in a
quiver diagram and the superpotential can be found by the
brane tiling \cite{BT1,BT2,BT3,BT4}.\footnote{See 
\cite{Kennaway,Yamazaki} for 
reviews of the quiver gauge theory and the brane tiling method.}
From these gauge theory data, we define a crystalline structure 
in three dimensions.
The crystal is composed of atoms of different colors, each
of which corresponds to a node of the quiver diagram and carries
a particular combination of D0 and D2 charges. The chemical bond 
is dictated by the arrows of the quiver diagram. There is a special
crystal configuration, whose exterior shape lines up with the toric 
diagram of the Calabi-Yau manifold. Such a crystal corresponds to 
a single D6 brane with no D0 and D2 charges. 
We define a rule to remove atoms from 
the crystal, which basically says that the crystal melts from its
peak. By using the non-commutative Donaldson-Thomas theory \cite{MR}, 
we show that there is a one-to-one correspondence between
molten crystal configurations and BPS bound states carrying 
non-zero D0 and D2 charges. The statistical model of crystal melting 
computes the index of D brane bound states. 

The number of BPS states depends on the choice of the stability condition,
and the BPS countings for different 
stability conditions are related to each other by
the wall crossing formulae. In this paper, we find that,
under a certain stability condition,
BPS bound states of D branes are counted by 
the non-commutative Donaldson-Thomas theory.
We can use the wall crossing formulae recently derived in
\cite{NN,Nagao} to relate this result to the commutative 
Donaldson-Thomas theory.
Since the topological string theory is
equivalent to the commutative Donaldson-Thomas theory 
for a general toric Calabi-Yau manifold \cite{MOOP},  
the relation \eqref{topstring} is indeed true for
some choice of the stability condition, as 
expected in \cite{INOV}. 

In general, the topological string partition function and the partition 
function of the  crystal melting model are not identical,
but their relation involves the wall crossing, 
\beq
   Z_{\rm crystal~melting} \sim Z_{\rm top}~~~({\rm modulo~wall~crossings}).
\eeq
This does not contradict with the result in  \cite{ORV, INOV} 
since there is no wall crossing phenomenon for $\bC^3$. 
In general, however, a proper understanding of the relation 
between the topological string theory and the crystal melting
requires that we take the wall crossing phenomena into account. 

In section 2, we will summarize the computation of 
D brane bound states from the gauge theory perspective. 
In section 3, we will discuss how this is related to
the recent mathematical results on the non-commutative
Donaldson-Thomas invariants. In section 4, we will 
formulate the statistical model of crystal melting for a general
toric Calabi-Yau manifold.  
The final section is devoted to summary of our result 
and discussion on the wall crossing phenomena.
The Appendix explains the equivalence of a configuration 
of molten crystal with a perfect matching of the bipartite graph.

\section{Quiver Quantum Mechanics}

In the classic paper by Douglas and Moore \cite{DM}, 
it was shown that the low energy effective theories
of D branes on some orbifolds are described by gauge theories 
associated to quiver diagrams. Subsequently, this result 
has been generalized to an arbitrary non-compact toric 
Calabi-Yau threefold. A toric Calabi-Yau threefold $X_{\Delta}$ 
is a fiber bundle of $T^2 \times \bR$ over $\bR^3$, where the
fibers are special Lagrangian submanifolds. The toric diagram 
$\Delta$ tells us where and how the fiber degenerates. For a given 
$X_\Delta$ and a set of D0 and D2 branes on $X_\Delta$, the following 
procedure determines the field content and superpotential
of the gauge theory on the branes. We will add a single D6 brane
to the system later in this section. 

\subsection{Quiver Diagram and Field Content}

The low energy gauge theory is a one-dimensional theory given 
by dimensional reduction 
of an ${\cal N}=1$ supersymmetric gauge theory in four 
dimensions. The field content of the theory is encoded in 
a quiver diagram, which is determined from the toric data 
and the set of D branes, as described in the following. 
A quiver diagram $Q=(Q_0,Q_1)$ consists of 
a set $Q_0$ of nodes, 
with a rank $N_i > 0$ associated to each node $i \in Q_0$, 
and a set $Q_1$ of arrows connecting the nodes. The corresponding 
gauge theory has a vector multiplet of gauge group $U(N_i)$ 
at each node $i$. There is also a chiral multiplet in the 
bifundamental representation associated to each arrow connecting
a pair of nodes. 

In the following, we will explain how to identify the 
quiver diagram. The reader may want to consult Figure 1,
which describes the procedure for the Suspended Pinched Point singularity, 
which is a Calabi-Yau manifold defined by the toric diagram
in Figure 1-(a) or equivalently by the equation,
\beq
   xy = z w^2,
\eeq
in $\bC^4$.

\begin{figure}[htbp]
\centering{\includegraphics[scale=0.4]{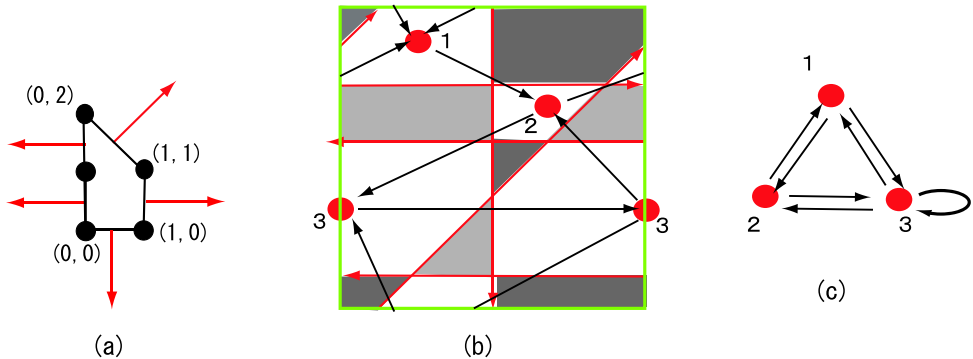}}
\caption{(a) The toric diagram for the Suspended Pinched Point
singularity.
(b) The configuration of D2 and NS5 branes after the T-duality on
$\bT^2$. The green exterior lines are periodically identified. 
The red lines representing NS5 branes 
separate the fundamental domain into several domains. 
The T-dual of D0 branes wrap the entire fundamental domain,
and fractional D2 branes are suspended between the red lines. 
The white domains contain D2 branes only. In each shaded
domain, there is an additional NS5 brane. There are two types 
of shades depending of the NS5 brane orientation. 
The white domains are connected
by arrows through the vertices, and the directions of the arrows are
determined by the orientation of the NS5 branes. (c) The quiver diagram
obtained by replacing the white domains of (b) by the nodes.}
\label{fig.SPPtiling}
\end{figure}

To identify the quiver diagram, we take T-dual of 
the toric Calabi-Yau manifold along the $\bT^2$ fibers \cite{BT2,FHKV}. 
The fibers degenerate at loci specified by the toric 
diagram $\Delta$, and the T-duality replaces the 
singular fibers by NS5 branes \cite{OV}. Some of these NS5 branes
divide $\bT^2$ into domains as shown in the red lines
in Figure 1-(b)
\cite{Imamura1,Imamura2,IIKY,Yamazaki}. 
The D0 branes become
D2 branes wrapping the whole $\bT^2$. The original
D2 are still D2 branes after the T-duality, but 
each of them is in a particular domain of $\bT^2$ 
suspended between NS5 branes. In addition, there are
some domains that contain NS5 branes stretched 
two-dimensionally in parallel with D2 branes.\footnote{
The NS5 branes are also filling the four-dimensional
spacetime $\bR^{1,3}$ while the D2 branes are localized along a timelike
path in four dimensions.} Let us denote the domains
without NS5 branes by $i \in Q_0$ and the domains with 
NS5 branes by $a \in I$. In Figure 1-(b), the $Q_0$-type
domains are shown in white, and the $I$-type domains are
shown with shade. There are two types of shades, corresponding
to two different orientations of NS5 branes. This distinction
will become relevant when we discuss the superpotential. 

The $Q_0$-type domains
are identified with nodes of the quiver diagram since 
open strings ending on them can contain massless excitations.
The rank $N_i$ of the node $i \in Q_0$ is the number of D2 branes
in the corresponding domain. On the other hand, $I$-type domains
give rise to the superpotential constraints as we shall see below.  
Though two domains $i, j \in Q_0$ never share an edge, they 
can touch each other at a vertex. In that case, open strings 
going between $i$ and $j$
contain massless modes. We draw an arrow from $i\rightarrow j$ 
or $i\leftarrow j$ depending on the orientation of the massless 
open string modes, which is determined by the orientation of 
NS5 branes.  Note that the quiver gauge theory 
we consider in this paper are in general chiral. 
This completes the specification of the quiver diagram. 

As another example,
the quiver diagram for the conifold geometry
has two nodes connected by two sets
of arrows in both directions. The ranks of the gauge groups
are $n_0$ and $n_0- n_2$, where $n_0$ and $n_2$ are the numbers
of D0 and D2 branes. The gauge theory is a dimensional reduction
of the Klebanov-Witten theory \cite{KW} when $n_2=0$ and 
the Klebanov-Strassler theory \cite{KS} when $n_2 >0 $.

\subsection{Superpotential and Brane Tiling}\label{tiling.sec}

Each domain $a \in I$ containing an NS5 brane 
is surrounded by domains $i_1, i_2, ..., i_n \in Q_0$ without NS5 branes, 
as in Figure 1-(b). By studying the geometry T-dual to $X_\Delta$ 
in more detail, one finds that the domain is contractible. Since 
open strings can end on the domains $i_1, i_2, ..., i_n$, 
the domain $a$ can give rise to worldsheet instanton
corrections to the superpotential. 

This fact, combined with the requirement that 
the moduli space of the gauge theory agrees with the geometric 
expectation for D branes on $X_\Delta$, determines the superpotential.
Depending on the NS5 brane orientation, 
the $I$-type domains are further classified into two types, $I_+$ and $I_-$,
and thus the regions of  torus is divided into three types $Q_0$, $I_+$ and $I_-$.
Such a brane configuration, or a classification of regions of $\bT^2$, is called the brane tiling.\footnote{In the literature the word brane tiling refers to the bipartite graph explained below.
Here the word brane tiling refers to a brane configuration  
as shown  Figure 1-(b). Such a graph is called the fivebrane diagram in \cite{IKY}.}
In Figure 1-(b), the brane tiling is shown by the two different shades. 
The superpotential $W$ is then given by
\beq
 W =  \sum_{a\in I_+} {\rm Tr}\left( \prod_{q=1}^{n_a^+}
A_{i^{(a)}_{q,+},i^{(a)}_{q+1,+}}\right)
   -   \sum_{a\in I_-} {\rm Tr}\left( \prod_{q=1}^{n_a^-}
A_{i^{(a)}_{q,-},j^{(a)}_{q+1,-}} \right), \label{eq.W}
\eeq
where the domain $a \in I_\pm$ are surrounded by 
the arrows $i^{(a)}_{1,\pm} \to i^{(a)}_{2,\pm} 
\to \cdots \to i^{(a)}_{n_a^\pm+1,\pm}
\to i^{(a)}_{1,\pm}$. For each arrow $i^{(a)}_{q,\pm} \to i^{(a)}_{q+1,\pm}$ 
($1 \leq q \leq n_a^\pm$), the corresponding bifundamental field is 
denoted by $A_{i^{(a)}_{q,\pm},i^{(a)}_{q+1,\pm}}$. 
This formula is tested in many examples. In particular, 
it has been shown that the formula reproduces the toric 
Calabi-Yau manifold $X_{\Delta}$ as the moduli space of the
quiver gauge theory \cite{FV}.

In the literature of brane tiling, bipartite graphs are often used
in place of brane configurations as in Figure 1-(b). 
A bipartite graph is a graph consisting of vertices colored 
either black or white and edges connecting black and white
vertices. Since bipartite graphs will also play roles in the following
sections, it would be useful to explain how it is related to our story
so far. For a given brane configuration, we can draw a bipartite graph 
on $\bT^2$ as follows. In each domain in $I_{+}$ ($I_-$), place a 
white (black) vertex. Draw a line connecting a white vertex 
in a domain $i\in I_{+}$ and a black vertex in a neighboring 
domain $j\in I_-$.  The resulting graph $\Gamma$ is bipartite.
See Figure 2 for the comparison of the brane configuration and
the bipartite graph in the case of the Suspended Pinched Point
singularity. We can turn this into a form that is more 
commonly found in the literature,
for example in \cite{BT2}, by choosing a different fundamental
region as in Figure 3.

\begin{figure}[htbp]
\centering{\includegraphics[scale=0.4]{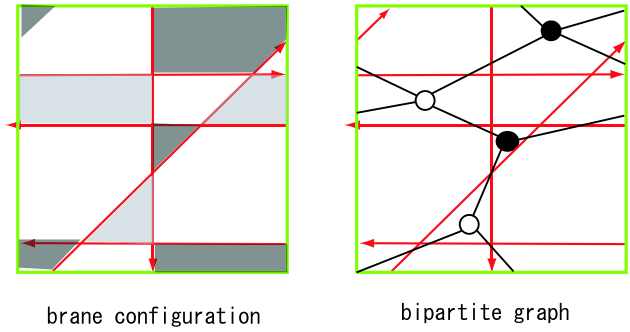}}
\caption{The correspondence between the brane configuration on 
$\bT^2$  and the bipartite graph. The white (black) vertex of the
bipartite graph corresponds to the region $I_+$ ($I_-$) in light (dark) 
shade. The edge of the bipartite graph corresponds to an
intersection of $I_-$ and in $I_+$. From this construction, it
automatically follows that the graph so obtained is bipartite.}
\label{fig.SPPbipartite}
\end{figure}

\begin{figure}[htbp]
\centering{\includegraphics[scale=0.3]{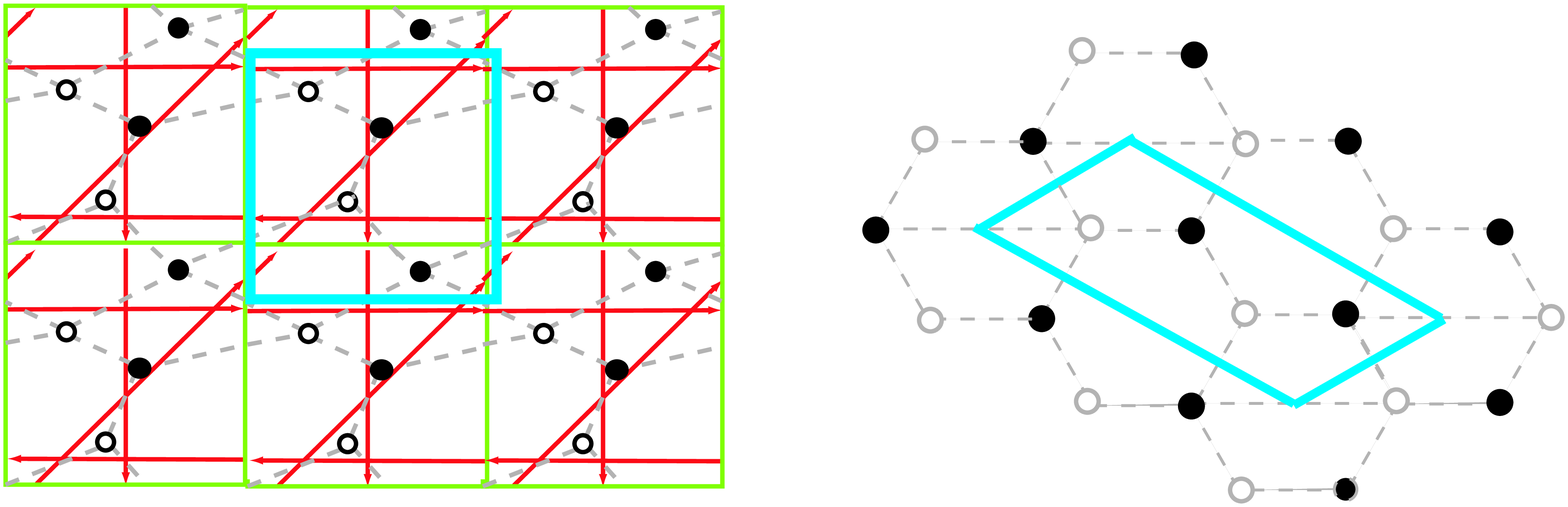}}
\caption{By choosing a different fundamental region
of $\bT^2$, we find a bipartite graph which is more commonly found in the literature.}
\label{fig.SPPregion}
\end{figure}

\subsection{D-term Constraints and the Moduli Space}

The F-term constraints are given by derivatives of the superpotential, 
which can be determined as in the above. 
The moduli space of solutions to the D-term constraints is
then described by a set of gauge invariant observables divided out by
the complexified gauge group $G_{\bC}$ \cite{LT}.  The theorem 
by King \cite{King} states that an orbit 
of $G_{\bC}$ contains a solution to the D-term conditions
if and only if we start with a point that satisfies the 
$\theta$-stability, a condition defined in the next section. 
Thus, we can think of the moduli space as 
a set of solutions to the F-term constraints obeying the
$\theta$-stability condition, modulo the action of $G_{\bC}$. 

\subsection{Adding a Single D6 Brane}

To make contact with the Donaldson-Thomas theory, 
we need to include one D6 brane.
Since the D6 brane fills the entire Calabi-Yau manifold, which is
non-compact, it behaves as a flavor brane. In the low energy
limit, the open string between the D6 brane and another D brane
gives rise to one chiral multiplet in the fundamental 
representation for the D brane on the other end.
The D6 brane then enlarges the quiver diagram by 
one node and one arrow from the new node. 
To understand why we only get one arrow
from the D6 brane, let us take T-duality along the $\bT^2$
fiber again. The D6 brane is mapped into a D4 brane which is a point 
in some region in $\bT^2$. This means that we only have one new 
arrow from the new node corresponding to the D6 brane to the
node corresponding to the D2 branes in the region. 
See \cite{Jafferis, Szabo} for related discussion in the literature.


\section{Non-commutative Donaldson-Thomas Theory}

In the previous section, we discussed how to construct
the moduli space of solutions to the F-term and D-term
constraints in the quiver gauge theory corresponding to
a toric Calabi-Yau manifold $X_\Delta$ with a set of D0/D2 branes 
and a single D6 brane. In this section, we will review
and interpret the mathematical formulation of the non-commutative
Donaldson-Thomas invariant in \cite{Szendroi, MR} for $X_\Delta$.
We find that it is identical to the Euler number 
of the gauge theory moduli space.

\subsection{Path Algebra and its Module}

For the purpose of this paper, modules are the same as representations. 
Consider a set of all open paths on the quiver diagram $Q=(Q_0, Q_1)$. By 
introducing a product as an operation to join a head of a path to a tail 
of another (the product is supposed to vanish if the head and the tail 
do not match on the same node) and by allowing formal sums of paths,
the set of open oriented paths can be made into an algebra $\bC Q$ called the 
path algebra. We would like to
point out that there is a one-to-one correspondence between 
a representation of the path algebra and a classical 
configuration of bifundamental fields of the quiver gauge theory. 
Suppose there is a representation $M$ of the path algebra. For each 
node  $i \in Q_0$, there is a trivial path $e_i$ of zero length that
begins and ends at $i$. It is a projection, $(e_i)^2 = e_i$. 
Since every path starts at some node $i$
and ends at some node $j$, 
the sum $\sum_i e_i$ acts as the identity on the path algebra. 
Therefore, 
$M = \oplus_{i\in Q_0} M_i$, where $M_i = e_i M$. 
Let us write $N_i = \dim M_i$. For each
path from $i$ to $j$, one can assign a map from $M_i$ to $M_j$.
In particular, there is an $N_i \times N_j$ matrix for each 
arrow $i \rightarrow j \in Q_1$ of the quiver diagram. By identifying this
matrix as the bifundamental field associated to the arrow
$i \to j$, we obtain a classical configuration 
of bifundamental fields with the gauge group $U(N_i)$ at 
the node $i$. By reversing the process, we can construct
a representation of the path algebra for each configuration of
the bifundamental fields. 

\subsection{F-term Constraints and Factor Algebra $A$} \label{F-term.subsec}

Let us turn to the F-term constraints. Since the bifundamental
fields of the quiver gauge theory is a representation of the
path algebra, the F-term equations give
relations among generators of the path algebra. It is natural to
consider the ideal ${\cal F}$ generated by the F-term equations
and define the factor algebra $A = \bC Q/{\cal F}$.
The bifundamental fields obeying the F-term constraints then
generate a representation of this factor algebra. Namely,
classical configurations of the quiver gauge theory 
obeying the F-term constraints are in one-to-one correspondence
with $A$-modules. 

As an example, the algebra $A$ for the conifold geometry contains an idempotent ring $\bC[e_1,e_2]$ generated by two elements and is given by the following four generators and relations:\footnote{The center $Z(A)$ of this algebra $A$ is generated by $x_{ij}=a_i b_j+b_j a_i (i,j=1,2)$, and is given by 
\beq
Z(A)=\bC[x_{11},x_{12},x_{21},x_{22}]/(x_{11}x_{22}-x_{12}x_{21}),
\eeq
which is the ring of functions of the conifold singularity.
}
\beq
A=\bC [e_1,e_2]\langle a_1,a_2,b_1,b_2 \rangle / 
\left(a_1 b_i a_2=a_2 b_i a_1, b_1 a_i b_2=b_2 a_i b_1  \right)_{i=1,2},
\eeq
Each $A$-module for this algebra corresponds 
to a choice of ranks of the gauge groups and a
configuration of the bifundamental fields $a_i, b_i$
satisfying the F-term constraints. 

\begin{figure}[htbp]
\centering{\includegraphics[scale=0.4]{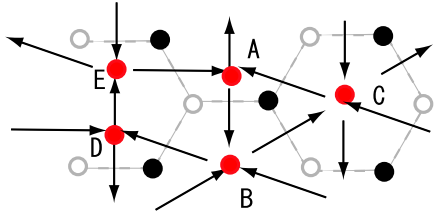}}
\caption{Representation of F-term constraints on the quiver diagram on $\bT^2$. In this example, if we write by $X_{AB}$ the bifundamental corresponding to an arrow starting from vertex $A$ and ending at $B$ etc., 
then the superpotential \ref{eq.W} contains a term
$
W=-\textrm{tr}(X_{AB}X_{BC}X_{CA})+\textrm{tr}(X_{AB}X_{BD}X_{DE}X_{EA}),
$
and the F-term condition for $X_{AB}$ (multiplied by $X_{AB}$) says that the product of bifundamentals fields along the triangle $ABC$ and that along the square $ABDE$ is the same.
}
\label{fig.SPPFterm}
\end{figure}

F-term constraints have a nice geometric interpretation on 
the quiver diagram, which we will find useful in the next section. 
We observe that each bifundamental field appears exactly twice 
in the superpotential with different signs of coefficients
in the superpotential shown in \eqref{eq.W}. 
By taking a derivative of the superpotential with respect
to a bifundamental field corresponding to a given arrow, 
the resulting F-term constraint states that 
the product of bifundamental fields around a face of 
the quiver on $\bT^2$ on one side of the arrow is equal to that 
around the face on the other side. See Figure \ref{fig.SPPFterm} 
for an example. Therefore, when we have a product of bifundamental
fields along a path, any loop on the path can be moved along
the path and the resulting product is F-term equivalent to the
original one.  In \cite{MR}, it is shown that for any point 
$i, j \in Q_0$, we can find a shortest path $v_{i,j}$ from $i$  to $j$ 
such that any other path $a$ from $i$ to $j$ is F-term equivalent to 
$v_{i,j} \omega^n$ with non-negative integer $n$, where $\omega$ 
is a loop around one face of the quiver diagram.
It does not matter where the loop $\omega$ is inserted along the 
path $v_{i,j}$ since different insertions are all F-term equivalent.
This means that any path is characterized by the integer $n$ and
the shortest path $v_{i,j}$. 

In the next subsection, we will impose the D-term constraints on
the space of finitely generated left $A$-modules, $\module A$. 
Before doing this, however, it is instructive to discuss 
topological aspects of $\module A$ by considering its bounded 
derived category\footnote{See \cite{Aspinwall} 
for an introductory explanation of derived categories 
in the context of string theory.} $D^b (\module A)$.
In mathematics, the algebra $A$ gives 
the so-called ``non-commutative crepant resolution'' \cite{VdB}.
For singular Calabi-Yau manifolds such as $X_\Delta$, 
the crepant resolution means a resolution that preserves
the Calabi-Yau condition.\footnote{Mathematically, we 
mean a resolution $f: Y_{\Delta}\to X_{\Delta}$ such 
that $\omega_Y=f^*\omega_X$, where $\omega_X$ and $\omega_Y$ 
are canonical bundles of $X$ and $Y$, respectively. 
For the class of toric Calabi-Yau threefolds, the existence 
of crepant resolution is known and different crepant resolutions 
 related by flops are equivalent in derived categories \cite{BridgelandFlop}.} 
For a crepant resolution $Y_\Delta$ of $X_\Delta$, we have the following
equivalence of categories\footnote{This is well-known in the case of 
the conifold (cf. \cite{VdB2}). For general toric Calabi-Yau threefolds, 
this is not yet mathematically proven as far as the authors are aware of, 
although there are proofs in several examples \cite{UY2,UY3,Nagao}.}:
\beq
D^b\left(\coh(Y_{\Delta})\right)\cong D^b (\module A),\label{eq.coh=mod}
\eeq
where $D^b(\coh (Y_{\Delta}))$ is a bounded derived category of coherent 
sheaves of crepant resolution $Y_{\Delta}$, 
and $D^b(\module A)$ is the bounded derived categories of 
finitely generated left $A$-modules.
The equation \eqref{eq.coh=mod} is also interesting from 
the physics viewpoint. 
Since $D^b(\coh (Y_{\Delta}))$ gives a topological classification 
of A branes on the resolved space $Y_\Delta$, the equivalence
means that $D^b (\module A)$ also classifies D branes, which 
is consistent with our interpretation above that 
$A$-modules are in one-to-one
correspondence with a configuration of bifundamental fields
obeying the F-term constraints. 

We should note that the paper \cite{MR}, which computes the non-commutative
Donaldson-Thomas invariants for general toric Calabi-Yau manifolds,
requires a set of conditions on brane tilings, namely on 
the superpotential. We find that most 
of their conditions (specified in lemma 3.5 and conditions 4.12) 
are automatically satisfied for any quiver gauge theories
for D branes on general toric Calabi-Yau manifold. We have not 
been able to prove that the condition 5.3 also holds in
general, but it is satisfied in all the examples we know.

\subsection{D-term Constraints and $\theta$-Stability}

We saw that the derived category $D^b (\module A)$ of $A$-modules
gives the topological classification of D branes in the toric Calabi-Yau
manifold $X_\Delta$. To understand the moduli space of D branes, however,
we also need to understand implications of the D-term constraints.
This is where the $\theta$-stability comes in.\footnote{The
$\theta$-stability is a special limit of $\Pi$-stability as 
discussed in \cite{DFR, Bridgeland}.}
Let $\theta\in \bN^{Q_0}$ be a vector whose components are real numbers. 
Consider an $A$-module $M$, and recall that this $M$ is decomposed 
as $M=\oplus_{i\in Q_0} M_i$ with $M_i = e_i M$. 
The module $M$ is called $\theta$-stable if
\beq
\sum_{i\in Q_0} \theta_i (\dim e_i M')> 0. \label{eq.thetastability}
\eeq
for every submodule $M'$ of $M$.\footnote{In some literature, 
an additional condition
$\sum_{i\in Q_0} \theta_i (\dim M_i)=0$ is imposed for a choice
of $\theta$. This is trivially satisfied for the choice $\theta=(0,0,\dots,0)$ we choose below.}
When $>$ is replaced by $\ge$, 
the module $M$ is called $\theta$-semistable. 

In the language of gauge theory, the stability 
condition \eqref{eq.thetastability} is required by the D-term conditions.
Some readers might wonder why the D-term conditions, which are equality
relations, can be replaced by an inequality as \eqref{eq.thetastability}. 
In fact, the similar story goes for the Hermitian Yang-Mills equations. 
There instead of solving the Donaldson-Uhlenbeck-Yau equations, 
we can consider holomorphic vector bundles with a suitable 
stability condition, the so-called $\mu$-stability or Mumford-Takemoto 
stability \cite{Donaldson,UY}. As we mentioned at the end of section 2, it is known that
a configuration of bifundamental fields is mapped to a solution 
to the D-term equations
by a complexified gauge transformation $G_{\bC}$ if and only if
the configuration is $\theta$-stable. 
Since each $A$-module $M$ gives a representation
of $G_{\bC} = \prod_{i \in Q_0} GL(N_i, \bC)$, where $GL(N_i, \bC)$
is represented by $M_i = e_i M$ at each node, each $A$-module
specifies a particular $G_\bC$ orbit. Thus, 
finding a $\theta$-stable module
is the same as solving the D-term conditions. 

Up to this point we have not specified the value of $\theta$. 
Physically, $\theta$'s correspond to the FI parameters, which 
are needed to write down D-term equations \cite{DFR}. 
Although the Euler number of the space of $\theta$-stable $A$-modules
does not change under infinitesimal deformation of $\theta$, 
it does change along the walls of marginal stability \cite{NN,Nagao}. 
The noncommutative Donaldson-Thomas invariant defined by 
\cite{Szendroi} is in a particular chamber in the 
space of $\theta$'s. Following \cite{MR}, we hereafter take 
$\theta=(0,0,\dots ,0)$. We will comment more about this issue 
in the final section.

\subsection{D6 Brane and Compactification of Moduli Space}\label{subsec.D6}

We have found that solutions to the F-term and D-term conditions
in the quiver gauge theory are identified with 
$\theta$-stable $A$-modules.
We want to understand the moduli space of such modules
and compute its Euler number. 

Since D brane charges correspond to the ranks of the gauge groups, 
we consider moduli space of $\theta$-stable modules with  
dimension $\dim M_i = N_i$ ($i \in Q_0$), which we denote by 
$\scM^{N}(A)$. To compute its Euler number, we need to address
the fact that the moduli space of stable $A$-modules is not always 
compact. In mathematics literature, the necessary compactification
is performed by enlarging the quiver diagram by adding
one more node in the following way. 

Let us fix an arbitrary vertex $i_0$, and define a new quiver 
$\hat{Q}=(\hat{Q}_0,\hat{Q}_1)$ by
\beq
\hat{Q}_0=Q_0 \cup \{*\}, \quad \hat{Q}_1=Q_1 \cup \{a_*:*\to i_0\}.
\eeq
Namely, we have added one new vertex $*$ and one arrow $* \to i_0$
to obtain the extended quiver diagram $\hat{Q}$. As in the previous
case for $Q$, we can define the path 
algebra $\bC \hat{Q}$, the ideal $\hat{\cal F}$ generated in $\bC Q$ 
by ${\cal F}$, and the factor algebra $\hat{A}=\bC \hat{Q}/\hat{{\cal F}}$. 
Define 
$\hat{\theta}\in \bN^{Q_0+1}$ by $\hat{\theta}=(\theta,1)$ 
and define $\hat{\theta}$-stable and semistable $\hat{A}$-modules 
using stability parameter $\hat{\theta}$. It is shown in lemma 2.3 of \cite{MR} 
that $\hat{\theta}$-semistable $\hat{A}$-modules are always 
 $\hat{\theta}$-stable, and the moduli space $\hat{\scM}_{i_0}^{N}(A)$ of
 $\hat{\theta}$-stable modules with specified dimension 
vector $\hat{N}\in \bN^{Q_0+1}$ is compact.

Adding the extra-node allows us to compactify the moduli space.
In the language of D branes, this corresponds to adding a
single D6 brane filling the entire Calabi-Yau manifold, which is 
necessary to interpret the whole system as a six-dimensional $U(1)$
gauge theory related to the Donaldson-Thomas theory. 
As we mentioned in section 2, the D6 brane serves
as a flavor brane and adds an extra node exactly in the way 
described in the above paragraph. 
Note that, in the above paragraph, the ideal $\hat{{\cal F}}$ is generated
by the original ideal ${\cal F}$. 
In the quiver gauge theory, this corresponds to
the fact that the flavor brane does not introduce a new gauge invariant
operator to modify the superpotential.
In this way, we arrive at the definition of non-commutative 
Donaldson-Thomas invariant as the Euler number $\chi
(\hat{\scM}_{i_0}^{N}(A))$
of cohomologies.
With our identification of $\hat{\scM}_{i_0}^{N}(A)$ with the moduli 
space of solutions to the F-term and D-term conditions, 
$\chi(\hat{\scM}_{i_0}^{N}(A))$ computes the Witten index of
bound states of D0 and D2 branes bound on a single D6 brane
(ignoring the trivial degrees of freedom corresponding the center of mass 
of D branes in $\bR^{1,3}$).

We have chosen a specific vertex $i_0$ to define the non-commutative 
Donaldson-Thomas invariant. The $i_0$ dependence drops out in simple 
cases such as $\mathbb{C}^3$ and conifold, but in general
$\chi(\hat{\scM}_{i_0}^N(A))$ depends on the choice of the $i_0$. We note that 
the quiver gauge theory discussed in section 2 also has an 
apparent dependence on $i_0$. There $i_0$ corresponds to the $Q_0$-type 
domain where the D6 brane is located after the T-duality. 
Since the fundamental group of the toric Calabi-Yau manifold $X_\Delta$ 
is trivial, there is no moduli intrinsic to the D6 brane before
taking the T-duality. Thus, we expect that the apparent $i_0$ dependence
in the gauge theory side should disappear with a proper treatment 
of the T-duality.  It would be interesting 
to study this point further.\footnote{In the example discussed in 
\cite{Nagao}, the result is dependent on the choice of $i_0$ and a 
particular vertex should be chosen in order to match with the category 
of perverse coherent sheaves. We thank Yukinobu Toda for discussions 
on this point.}

\section{Crystal Melting}

In this section, we define a statistical mechanical model
of crystal melting and show that the model reproduces the 
counting of BPS bound state of D branes. Using
the quiver diagram and the superpotential of the gauge theory,
we define a natural crystalline structure 
in three dimensions.
We specify a rule to remove atoms from the crystal
and show that each molten crystal corresponds to a particular
BPS bound states of D branes. We use the result of 
\cite{MR} to show that all the relevant BPS states are 
counted in this way. 

\subsection{Crystalline Structure}\label{crystal.subsec}

Mathematically, the three-dimensional crystal we define here is
equivalent to a set of basis for $A e_{i_0}$, where $A$ is the factor algebra
$A = \bC Q/{\cal F}$ of the path algebra $\bC Q$ divided by the 
ideal ${\cal F}$ generated by the F-term constraints and
$e_{i_0}$ is the path of zero length at the reference node $i_0$,
which is also the projection operator to the space of paths
starting at $i_0$. Colloquially, the crystal is a set of paths
starting at $i_0$ modulo the F-term relations. As we shall see 
later, it corresponds to a BPS state corresponding to a single 
D6 brane with no D0 and D2 charges. 
We interpret $A e_{i_0}$ in terms of a three-dimensional crystal as follows. 

The crystal is composed of atoms piled up 
on nodes in the universal covering $\tilde{Q}$ on $\bR^2$. By using 
the projection, $\pi: \tilde{Q} \to Q$, each atom is assigned with 
a color corresponding to the node in the original quiver diagram $Q$. 
The arrows of the quiver diagram determines the chemical bond between 
atoms. We start by putting one atom on the top of the reference node 
$i_0$. Next attach an atom at an adjacent node $j\in \tilde{Q}_0$ that is
connected $i_0$ by an arrow going from $i_0$ to $j$. The atoms at such nodes 
are placed lower then the atom at $i_0$. In the next step, start with 
the atoms we just placed and follow arrows emanating from them
to attach more atoms at the heads of the arrows. 

As we repeat this procedure, we may return back to a node where 
an atom is already placed. In such a case, 
we use the following rule. As we explained in
section 3.2, modulo F-term constraints,
any oriented path $a$ starting at $i_0$ and ending at $j$ can 
be expressed as $v_{i_0,j}\omega^n $, where $\omega$ is 
the loop around a face in the quiver
diagram 
 and $v_{i_0,j}$ is one of the shortest
paths from $i_0$ to $j$. This defines an integer $h(a)=n$
for each path $a$. The rule of placing atoms is that, 
if a path $a$ takes $i_0$ to $j$ and if $h(a)=n$, 
we place an atom at the $n$-th place under the first atom 
on the node $j$. If there is already an atom at the 
$n$-th place, we do not place a new atom.

By repeating this procedure, we continue to attach atoms
and construct a pyramid consisting of infinitely many atoms.
Since atoms are placed following paths from $i_0$ modulo 
the F-term relations, it is clear that atoms in the crystal 
are in one-to-one correspondence with basis elements of $A e_{i_0}$. 
Note that by construction the crystal has a single peak at the 
reference node $i_0$. 

\begin{figure}[htbp]
\centering{\includegraphics[scale=0.2]{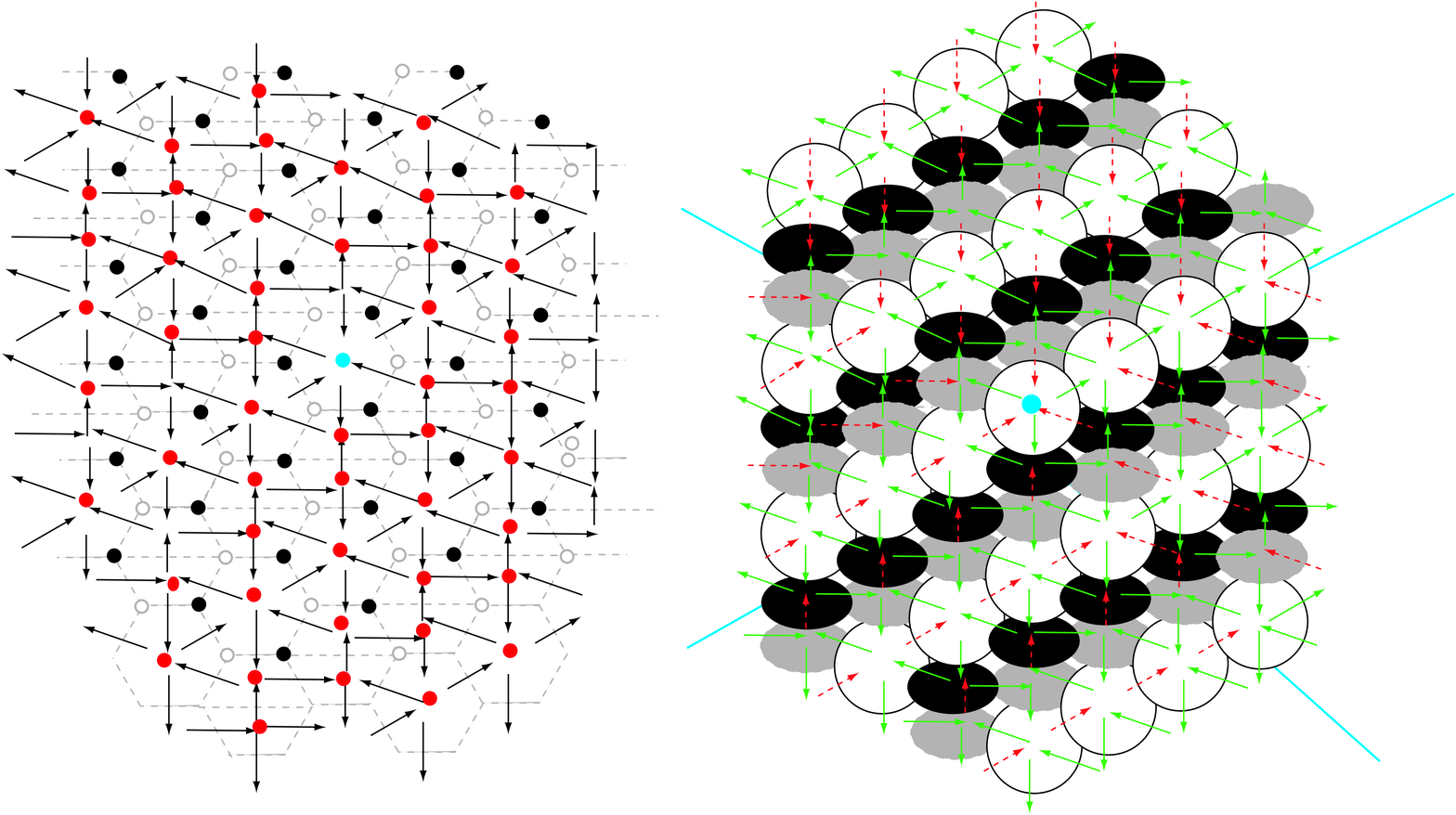}}
\caption{Starting from the universal cover $\tilde{Q}$ of quiver $Q$ shown on the left, we can construct a crystal on the right. 
Each atom carries a color
corresponding to a node in $Q$, and they are connected by
arrows in $\tilde{Q}_1$. The green arrows represent arrows on the surfaces of the crystal, whereas the red ones are not.
In the case of the Suspended Pinched Point singularity, 
the atoms come with 3 colors (white, black and gray),
corresponding to the 3 nodes of the original quiver diagram $Q$ on $\bT^2$ shown in Figure \ref{fig.SPPtiling}.}
\label{fig.SPPatom}
\end{figure}

This defines a crystalline structure
for an arbitrary toric Calabi-Yau manifold. 
In particular, it reproduces the crystal for $\bC^3$
discussed in \cite{ORV,INOV}, and the one for conifold 
in \cite{Szendroi}.  See Figure \ref{fig.SPPatom}
for the crystalline structure corresponding to
the Suspended Pinched Point singularity. In this example, 
the ridge of the crystal (shown as blue lines in 
Figure \ref{fig.SPPatom}) coincides with the $(p,q)$-web of 
the toric geometry. As we will discuss later, this is a 
general property of our crystal.

\subsection{BPS State and Molten Crystal}\label{molten.subsec}

In the forthcoming discussions, the crystal defined above will be 
identified with a single D6 brane with no D0 and D2 charges. Bound
states with non-zero D0 and D2 charges are obtained by removing
atoms following the rule specified below. 

In \cite{Szendroi,MR}, the 
Donaldson-Thomas invariants $\chi(\hat{\scM}_{i_0}^{N}(A))$
are computed by using the $U(1)^{\otimes 2}$ symmetry of the 
moduli space  $\hat{\scM}_{i_0}^N$ corresponding to translational
invariance of $\bT^2$. By the standard localization techniques, 
the Euler number can be evaluated at the fixed point
set of the moduli space under the symmetry. Correspondingly, 
in the gauge theory side, BPS states counted by the index are those
that are invariant under the global $U(1)^{\otimes 2}$ symmetry acting on 
bifundamental fields preserving the F-term constraints since those
do not have extra zero modes and do not contribute to the index. 
We are interested in counting such BPS states. 

In order for a molten crystal to correspond to $U(1)^{\otimes 2}$ 
invariant $\hat{\theta}$-stable $A$-modules, we need to impose 
the following rule to remove atoms from the crystal. Let $\Omega$
be a finite set of atoms to be removed from the crystal. 

\medskip
\noindent
{\bf The Melting Rule}: 
If $a\alpha$ is in $\Omega$ for some $a \in A$, 
then $\alpha$ should also be in $\Omega$. 

\begin{figure}[htbp]
\centering{\includegraphics[scale=0.2]{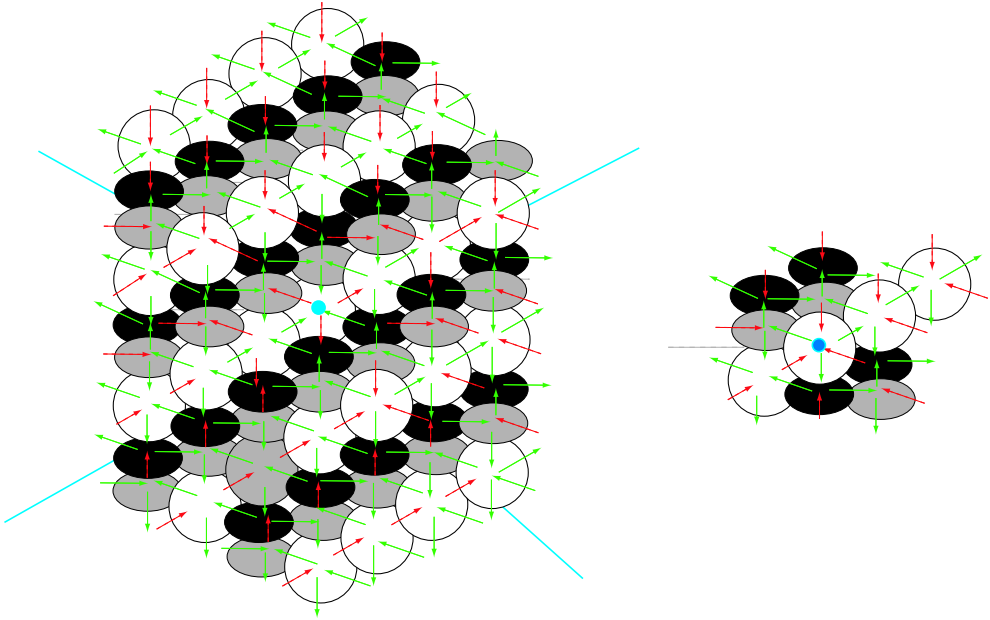}}
\caption{Example of a molten crystal and its complement $\Omega$. In this example $\Omega$ contains 12 atoms, one hidden behind an atom on the reference point represented by a blue point.
It is easy to check that $\Omega$ satisfies the melting rule mentioned in the text.}
\label{fig.SPPmolten}
\end{figure}

\medskip
\noindent
Since atoms of the crystal correspond to elements of $A e_{i_0}$, 
we used the natural action of $A$ on $A e_{i_0}$ to define $a\alpha$ in
the above. This means that crystal melting starts at the peak at $i_0$
and takes place following paths in $A e_{i_0}$. 
An example of a molten crystal satisfying 
this condition is shown in Figure \ref{fig.SPPmolten}.

The melting rule means that a complement ${\cal I}$ 
of the vector space spanned by $\Omega$ in $A e_{i_0}$ gives
an ideal of $A$. To see this, we just need to take the contraposition of 
the melting rule. It states:
For any $\beta \in {\cal I}$ and
for any $a \in A$, $a \beta$ is also in ${\cal I}$. 

\medskip
Generally speaking, an ideal of an algebra defines a module.
To see this, consider a vector $|{\cal I} \rangle$ which is 
annihilated by all elements of the ideal ${\cal I}$. 
From $|{\cal I} \rangle$,
we can generate a finite dimensional
representation of the algebra $A$ by acting elements 
of $A$ on it. However, the converse is not always true. 
Fortunately, when modules are $\hat{\theta}$-stable
and invariant under the 
$U(1)^{\otimes 2}$ symmetry, it was shown in \cite{Szendroi,MR} that 
there is a one-to-one correspondence between ideals and modules.
It follows that our molten crystal configurations are
also in one-to-one correspondence with $A$-modules and therefore with
relevant BPS bound states of D branes. This proves that the statistical 
model of crystal melting computes the index of D brane bound states.

It would be instructive to understand explicitly 
how each molten crystal configuration corresponds to a BPS bound state.  
Starting from a molten crystal specified by $\Omega$,
Prepare a one-dimensional vector space $V_{\alpha}$  
with basis vector $e_{\alpha}$ for each atom $\alpha\in \Omega$. 
For each arrow $a$ of $\tilde{Q}$, define the action of $a$ on $V_{\alpha}$ by
$a(e_{\alpha})=e_{\beta}$ when the arrow $a$ begins from $\alpha$ and ends an another atom $\beta \in \Omega$. Otherwise $a(e_{\alpha})$ is defined to be zero. 
Since an arbitrary path is generated by concatenation of arrows, we have defined an action of $a\in A$ on each $V_{\alpha}$. By linearly extending the action of $a$ onto the total space $M=\oplus_{\alpha\in \Omega} V_{\alpha}$, we obtain
a $A$-module $M$. 

There are several special properties about this module $M$. 
First, the F-term relations are automatically
satisfied. This is because when there exists two different paths $a,b\in A$ 
starting at $\alpha$ and ending at $\beta$, 
$a(e_{\alpha})$ and $b(e_{\alpha})$ are both defined to be $e_{\beta}$.
Second, by construction $M$ is generated by action of 
the algebra $A$ on 
a single element $e_{i,0}\in V_{i,0}$. 
In such a case $M$ is called a cyclic $A$-module, and by lemma 2.3 of \cite{MR} is also $\hat{\theta}$-stable. 
Third, by the cyclicity of the module it follows that $M$ is $U(1)^{\otimes 2}$ invariant up to gauge transformations.
Therefore, $M$ is a $U(1)^{\otimes 2}$ invariant 
$\hat{\theta}$-stable module. It follows from the 
result of \cite{MR} discussed at the beginning of this section 
that $M$ indeed corresponds a bound state of D branes 
contributing to the Witten index.

At the beginning of this subsection, we stated without 
explanation that the original crystal corresponds to 
a single D6 brane with no D0 and D2 charges, and removing
atoms correspond to adding the D brane charges. To understand
this statement, let us recall that, in section 2, 
we started with a configuration of D0 and D2 branes on 
the toric Calabi-Yau manifold and took a T-duality along the fiber
to arrive at the brane configuration. Thus, the number of D2 branes
at each node $j$ of the quiver diagram $Q$ is a combination of D0 and D2
charges before T-duality. It is this number that 
is equal to the rank of the gauge group at $j$.

By using the projection $\pi: \tilde{Q}_o \rightarrow Q_0$, 
the $A$-module $M$ is decomposed as 
$M=\oplus_{j \in Q_0} M_j$ as we saw in section 3, where
\beq
M_j=\bigoplus_{\alpha \in \Omega, \pi(\alpha)=j } 
V_{\alpha}\ . \label{backtomodule}
\eeq
In particular, the formula \eqref{backtomodule} means that 
the rank of the gauge group $N_j = {\rm dim} M_j$ at the node $j$
is equal to the number of atoms with the color corresponding
to the node $j$ that have been removed from the crystal. 
Thus, removing an atom at the node $j$ is equivalent to 
adding D0 and D2 charges carried by the node $j$. 
It is interesting to note that each atom in the crystal does not correspond
to a single D0 brane or a single D2 brane, but each of them carries
a specific combination of D0 and D2 charges. In the crystal melting picture,
fundamental constituents are not D0 and D2 branes but the atoms. 
This reminds us of
the quark model of Gell-Mann and Zweig, where the fundamental
constituents carry combinations of quantum numbers of hadrons, as
opposed to the Sakata model, where existing elementary particles
such as the proton, neutron and $\Lambda$ particle are
chosen as fundamental constituents.

\subsection{Observations on the Crystal Melting Model}

We would like to make a few observations on the statistical model 
of crystal melting that counts the number of BPS bound states of D branes. 

We have studied several examples of toric Calabi-Yau manifolds and
found that the crystal structure in each case matches with 
the toric diagram. In 
particular, the ridges of the crystal, when projected onto the 
$\bR^2$ plane, line up with the $(p,q)$ web of the maximally degenerate
toric diagram. This phenomenon is discussed in Appendix. There, we
also explain the correspondence between molten crystal configurations
and perfect matching of the bipartite graph introduced in section 2.2.

So far, we have considered molten crystals that are obtained
by removing a few atoms. We may call them low temperature
configurations. The high temperature behavior of the model, 
describing bound states with large D0 and D2 charges, 
is also interesting. For $\bC^3$, it was shown in \cite{ORV,INOV} 
that the high temperature limit of
the crystal melting model reproduces the geometric shape of the mirror 
manifold. Since the high temperature limit of a general statistical
model of random perfecting matchings is known to be described by a
certain plane algebraic curve \cite{KOS}, it would be interesting to 
understand its relation to the mirror of a general toric Calabi-Yau manifold
\cite{OY}.

In the last subsection, we found it useful to describe BPS bound
states using ideals of the algebra $A$. In the case when the toric
Calabi-Yau manifold is $\bC^3$, 
ideals are closely related to  the quantization of the
toric structure as discussed in \cite{INOV}. The
gauge theory for $\bC^3$ 
is the dimensional reduction of the ${\cal N}=4$ 
supersymmetric Yang-Mills theory
in four dimensions down to one dimension, and the
bifundamental fields are three adjoint fields. The F-term and D-term conditions
require that they all commute with each other. Thus chiral ring is generated 
by three elements $x, y, z$ which commute with each other without any further 
relation. In this case, any ideal ${\cal I}_\pi$ is characterized by 
the three-dimensional Young diagram $\pi$. 
Locate each box in the 3d Young diagram 
$\pi$ by the Cartesian coordinates $(i,j,k)$ ($i,j,k = 1, 2, 3, ...$)
of the corner of the box most distant
from the origin, and define $\Omega_\pi$ to be a set of 
the 3d Cartesian coordinates $(i,j,k)$ for boxes in $\pi$. We can then
define the ideal $\omega_\pi$ of the chiral ring by,
\beq
{\cal I}_\pi = \{ x^{i-1} y^{j-1} w^{k-1} | (i,j,k) 
\notin \Omega_\pi
\}. \label{ideal}
\eeq
In \cite{INOV}, this description was obtained
by quantizing the toric geometry by using its canonical K\"ahler form
and by identifying $x^{i-1} y^{j-1} w^{k-1}$ as states in the Hilbert
space. 

This can be generalized to an arbitrary toric Calabi-Yau manifold
$X_\Delta$ as follows. One starts with the quiver diagram corresponding to
$X_\Delta$ and use the brane tiling to identify the F-term equations. 
This gives the chiral ring generated by bifundamental fields obeying
the F-term and D-term relations. As we saw in section 4.3, each BPS
bound state is related to an ideal of the chiral algebra. 
We expect that such ideals arise from quantization of the toric structure.
BPS bound states of D branes emerging from the quantization of 
background geometry is reminiscent of the bubbling AdS space of 
\cite{LLM} and Mathur's conjecture on black hole microstates
\cite{Mathur:2005zp}.

\section{Summary and Discussion}\label{sec.conc}

In this paper, we established the connection between the counting
of BPS bound states of D0 and D2 branes on a single D6 brane to
the non-commutative Donaldson-Thomas theory. We studied the moduli space 
of solutions to the F-term and D-term constraints of the quiver
gauge theory which arises as the low energy limit of the brane
configuration. We found the direct correspondence between
the gauge theory moduli space and the space of modules of the factor 
algebra of the path algebra for the quiver diagram quotiented
by its ideal related to the F-term constraints, subject
to a stability condition to enforce the D-term constraints. 
Using this correspondence, we found
a new description of BPS bound states
of the D branes in terms of the statistical model of crystal
melting. The crystalline structure is determined by
the quiver diagram and the brane tiling which characterize 
the low energy effective theory of D branes. The crystal 
is composed of atoms of different colors, each of which 
corresponds to a node of the quiver diagram, and the chemical 
bond is dictated by the arrows of the quiver diagram. BPS states 
are constructed by removing atoms from the crystal.

The relation between the commutative and non-commutative Donaldson-Thomas
invariant has been extensively discussed in the recent literature. 
The degeneracy of D brane bound states changes when the value of 
$\theta$, used to define the stability condition, jumps along the 
codimension one subspace, which is called walls of marginal stability. 
The jump in the degeneracy can be computed by the wall crossing formula 
\cite{DenefM,KontsevichS,GMN}, and if we start from a particular chamber and 
applying wall crossing formula, we can obtain the value of 
$\chi(\hat{\scM}_i^{N}(A))$ in any chamber we want. 
In the example of conifold 
\cite{NN}, wall crossing relates non-commutative Donaldson-Thomas 
invariants to commutative Donaldson-Thomas invariants and to new invariants defined by Pandharipande and Thomas \cite{PT}. 
This story is further generalized by \cite{Nagao} when $\Delta$ has 
no internal lattice point. When the toric diagram contains an internal 
lattice point, non-commutative Donaldson-Thomas invariant includes 
D4 branes, since $H_4(Y_{\Delta})\ne 0$.  Since (commutative) 
Donaldson-Thomas invariants does not include D4 brane 
charges, the above discussion of wall crossing should be modified.

It has been proven recently that the topological 
string theory is  equivalent to the commutative Donaldson-Thomas theory 
for a general toric Calabi-Yau manifold \cite{INOV,MOOP}.
Since the commutative Donaldson-Thomas theory
count BPS states for some choice of stability
condition, 
\beq
Z_{\rm BH} = Z_{\rm top},
\eeq
is indeed true in some chamber. 
On the other
hand, our result shows that the relation,
\beq
    Z'_{\rm BH} = Z_{\rm crystal~melting},
\eeq
holds in another chamber, where $Z'_{\rm BH}$ is the BPS state counting
for another choice of the stability condition. 
Combining these two results, we find that 
the topological string theory and the 
statistical model of crystal melting 
are related by the wall crossing, and we have
\beq
   Z_{\rm crystal~melting} \sim Z_{\rm top}~~~({\rm modulo~wall~crossings}).
\eeq
Since there is no wall crossing phenomenon for the Donaldson-Thomas
theory on $\bC^3$, this result does not contradict with \cite{INOV},
where a direct identification of the topological string theory
and the crystal melting is made for $\bC^3$.
In general, we expect that a proper understanding of the relation 
between the topological string theory and the crystal melting
requires that we take the wall crossing phenomena into account.

The OSV formula (1.1) suggests yet another relation between the black 
hole microstate counting and the topological string theory. According
to \cite{GSY}, for a compact Calabi-Yau manifold, the D6/D2/D0 brane system 
gives rise to a large black hole in four dimensions since it is 
related to a spinning M theory black hole by the Kaluza-Klein reduction. 
In fact, one can compute the semi-classical Bekenstein-Hawking entropy 
for such a 4-dimensional black hole and find that it can be made arbitrarily 
large provided the D2 charges are sufficiently larger than the D0 charge. 
In this paper, we discussed the D6/D2/D0 system on a non-compact toric 
Calabi-Yau manifold with an infinite volume. Though it is not obvious 
that the gravity description in four dimensions is applicable in this
case, the OSV formula has been successfully tested
for a similar class of non-compact Calabi-Yau manifolds
\cite{Vafatorus,NoncompactOne,NoncompactTwo,NoncompactThree}.
If it is applicable in our case, it would imply the relation between 
$ Z_{\rm crystal~melting}$ and $|Z_{\rm top}|^2$, modulo wall crossings. It would be interesting
to find out if such a relation holds.

It appears that the crystal melting picture is closely related to the
quantization of the toric structure of the Calabi-Yau manifold. It would
be interesting to understand the relation better. This could lead to a
new insight into quantum geometry, along with the observations in
\cite{LLM} for the bubbling AdS geometry and \cite{Mathur:2005zp}
for black hole microstates. 

\section*{Acknowledgments}

We would like to thank Kentaro Nagao, Kazutoshi Ohta, Yukinobu Toda, Kazushi Ueda and Xi Yin for discussions. This work is supported in part by DOE grant DE-FG03-92-ER40701 and by the 
World Premier International Research Center Initiative of MEXT of Japan.
H.~O. is also supported in part by a Grant-in-Aid for Scientific 
Research (C) 20540256 
of JSPS and by the Kavli Foundation. M.~Y. is also supported in part 
by the JSPS fellowships for Young Scientists and by the 
Global COE Program for Physical Sciences Frontier at 
the University of Tokyo funded by MEXT of Japan.

\appendix

\section{Perfect Matchings}
In this Appendix we are going to explain the one-to-one correspondence between a molten crystal discussed in the main text and a perfect matching of the bipartite graph. 
This means that the problem of counting BPS states can also be reformulated as a problem of counting  perfect matchings of the bipartite graph, where a perfect matching is a subset of edges of the bipartite graph such that each vertex is contained exactly once.
The contents of this appendix is basically a recapitulation of \cite{MR}.

In section \ref{crystal.subsec} we considered a quiver $\tilde{Q}=(\tilde{Q}_0,\tilde{Q}_1)$, which is a universal cover of the quiver $Q$ on $\bT^2$. The dual graph of $\tilde{Q}$, which we denote by $\tilde{\Gamma}$, can be made bipartite using 
orientation of arrows of $\tilde{Q}$, and is a universal cover of 
the bipartite graph $\Gamma$ on $\bT^2$ described in section 2.2.
What we are going to do is to give an explicit correspondence between a perfect matching of the bipartite graph 
$\tilde{\Gamma}$ and a configuration of molten crystal.

We first construct a perfect matching from a molten crystal.
Given a molten crystal as shown in Figure \ref{fig.SPPatom}, choose all the arrows of $\tilde{Q}$ which are along the surface of the crystal. In the example of Figure \ref{fig.SPPatom}, 
such arrows are colored green in Figure \ref{fig.SPPatom}, 
while the remaining arrows are colored red. 
Take the set of the dual of edges colored red. 
It is proven by \cite{MR} that such a subset of edges of $\tilde{\Gamma}$ is a perfect matching. This is the perfect matching we wanted to construct.

\begin{figure}[htbp]
\includegraphics[scale=0.2]{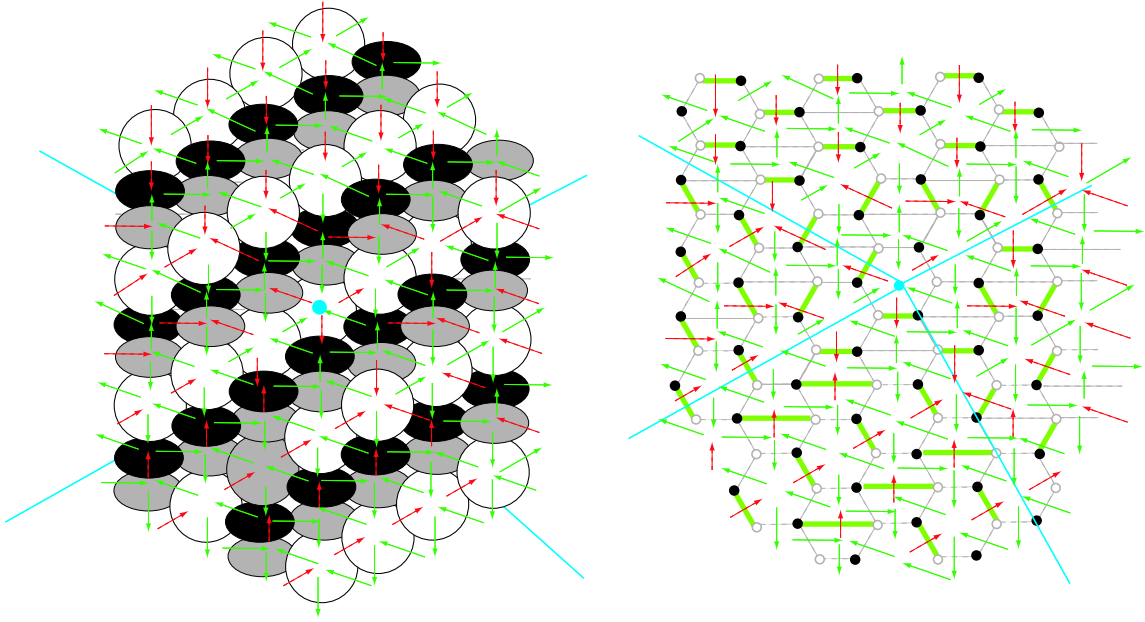}
\caption{Given a configuration of a molten crystal, we can construct a perfect matching of the bipartite graph. Each arrow is colored green if it is along the surface of the crystal, and red otherwise. The set of dual of arrows colored red gives a perfect matching of the bipartite graph.}
\label{fig.SPPmodule}
\end{figure}

In the case when no atoms are removed from the crystal, the perfect matching obtained by this method is called the canonical perfect matching, which we denote by $D_0$. Since 
only a finite number of atoms are removed from the crystal, 
the perfect matching obtained from a molten crystal by the above method coincides with $D$ when sufficiently away from the reference point $i$.

\begin{figure}[htbp]
\centering{\includegraphics[scale=0.2]{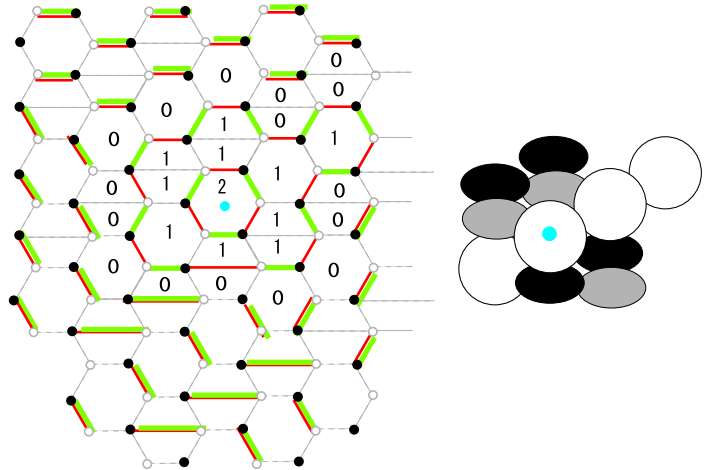}}
\caption{By superimposing a perfect matching of Figure \ref{fig.SPPmodule} with the canonical perfect matching shown later in Figure \ref{fig.SPPempty}, we have a  set of loops, which defines a height function $h_D$. From this function we can recover a molten crystal.}
\label{fig.SPPheight}
\end{figure}

Conversely, given a perfect matching $D$ which coincides with $D_0$ when sufficiently away from the reference point $i$, we can reproduce a molten crystal.
Let us superimpose $D$ with $D_0$, and we have a finite number of loops, as shown in Figure \ref{fig.SPPheight} in the case of Suspended Pinched Point. 
Define a height function $h_D$ such that 

\medskip
(1) $h_D(j)=0$ when sufficiently away from $i$.

\smallskip
(2) $h_D$ increases by one whenever we cross the loop and go inside it.
\medskip

\noindent The example of $h_D$ for the case of Suspended Pinched Point is shown in Figure \ref{fig.SPPheight}.

By removing $h_D(j)$ atoms from each $j\in \tilde{Q}_0$, we can construct a molten crystal. 
It was proven in \cite{MR} that the set of atoms removed from the crystal so defined satisfies the melting rule of section \ref{molten.subsec}. 
This establishes the one-to-one correspondence between a molten crystal and a perfect matching of the bipartite graph, meaning that BPS states can also be counted by perfect matchings of the bipartite graph $\tilde{\Gamma}$.

\begin{figure}[htbp]
\includegraphics[scale=0.2]{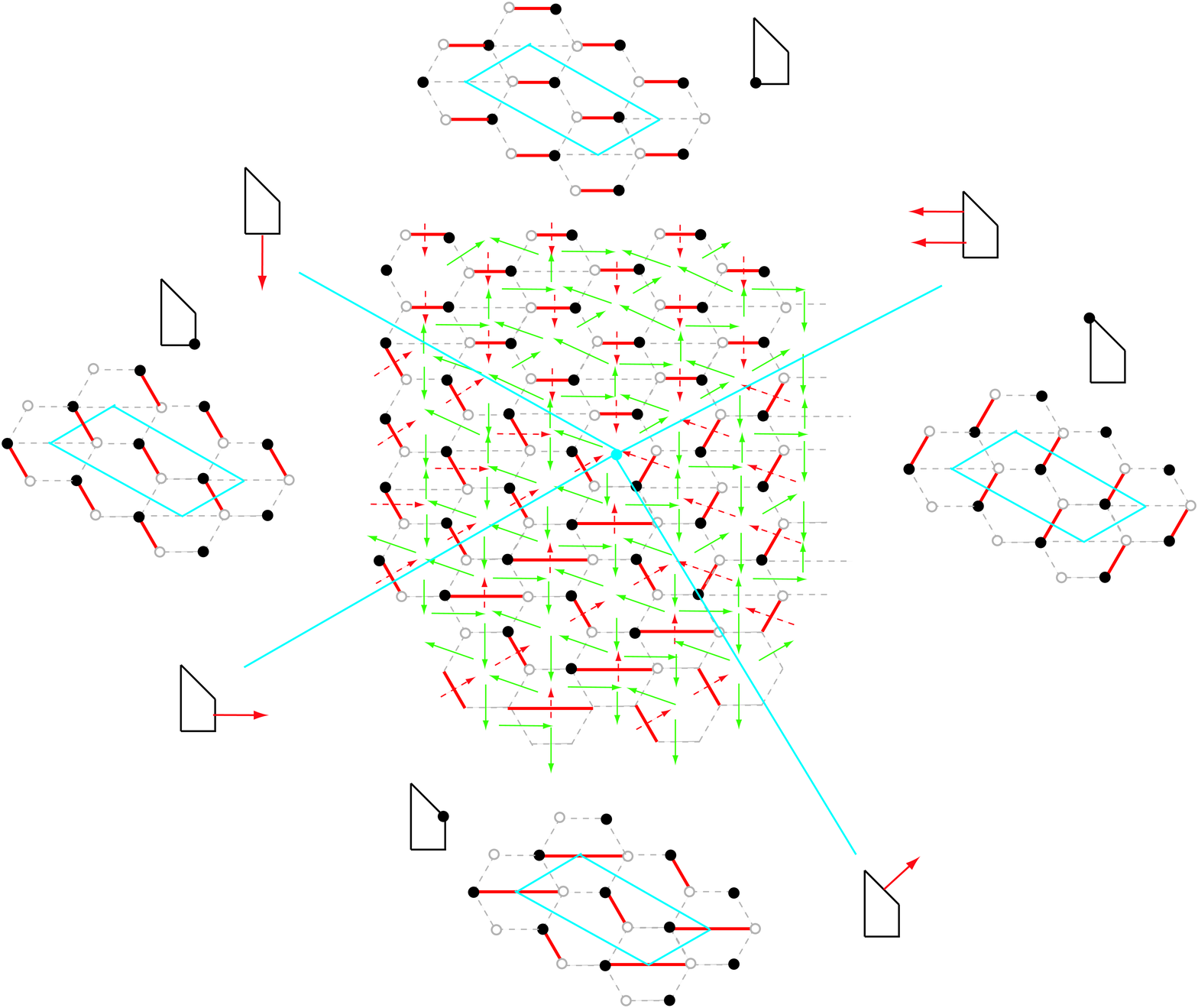}
\caption{The canonical perfect matching of the bipartite graph for
the Suspended Pinched Point singularity. Asymptotically, 
the perfect matching corresponds to one of the
 four perfect matching of the bipartite graph corresponding 
to vertices of the toric diagram. The blue borders between 
different choices of perfect matchings
represents the $(p,q)$-web.}
\label{fig.SPPempty}
\end{figure}

Finally, let us finish this Appendix by pointing out  an interesting connection
 of the canonical perfect matching $D_0$  with toric geometry.
The example of 
 canonical perfect matching $D_0$ for the Suspended Pinched Point is shown in Figure \ref{fig.SPPempty}. 
In this example, the asymptotic form of the bipartite graph 
has four different patterns. 
Each of four patterns is periodic and therefore an be thought of as a perfect matching of the bipartite graph on $\bT^2$.
 In the brane tiling literature, 
a perfect matching on the bipartite graph on $\bT^2$ is 
known to correspond to one of the lattice points of the 
toric diagram \cite{BT1,FV}.\footnote{For this correspondence, 
we consider superimposition of perfect matchings and 
define a $\bZ^2$-valued height function, which is similar
to the height function $h_D$ defined previously.}
We recognize that
the four perfect matchings are identified with the four corners
of the toric diagram in Figure 1-(a) and that 
the borders between different patters are identified with 
the blue lines in Figure 1-(b), which makes 
the $(p,q)$-web of the diagram.

In general, for an arbitrary toric Calabi-Yau manifold, we can use the same 
pattern to construct a perfect matching.
Divide the universal covering of the bipartite graph
into segments separated by the $(p,q)$-web of the toric 
diagram.\footnote{We choose the diagram that corresponds to 
the most singular Calabi-Yau manifold.}
The perfect matching in each segment is periodic and 
is identified with one of the perfect matchings of bipartite 
graphs on $\bT^2$, which corresponds to one of the lattice points of the 
toric diagram and 
the lattice point in question is precisely the vertex 
surrounded by the two $(p,q)$-webs on $\bT^2$.\footnote{In a consistent quiver gauge theory, 
it is believed that the multiplicity of perfect matchings 
at the vertices of the toric diagram is one \cite{BT4}.}
This determines a perfect matching. In particular,
this means that the ridges of the crystal line up with the
$(p,q)$-web of the toric diagram. 

We have examined several other examples as well, and
this pattern holds in all cases. Thus, we conjecture that
the perfect matching constructed in this way is canonical. 
We would like to stress again that this conjecture is not
needed to construct the crystal melting model. 
Here we are simply pointing out that, in the examples
we have studied, the crystalline structures fit beautifully 
with the corresponding toric geometries.


\end{document}